\documentclass[aps,prd,superscriptaddress,preprint,tightenlines,nofootinbib]{revtex4}

\usepackage{graphicx}
\usepackage{dcolumn}
\usepackage{bm}

\newcommand{\text}{\mathrm}

\begin{document}

\preprint{CLNS 09/2052}
\preprint{CLEO 09-03}

\title{\boldmath Inclusive Hadron Yields from $D^{+}_{s}$ Decays}

\author{S.~Dobbs}
\author{Z.~Metreveli}
\author{K.~K.~Seth}
\author{B.~J.~Y.~Tan}
\author{A.~Tomaradze}
\affiliation{Northwestern University, Evanston, Illinois 60208, USA}
\author{J.~Libby}
\author{L.~Martin}
\author{A.~Powell}
\author{C.~Thomas}
\author{G.~Wilkinson}
\affiliation{University of Oxford, Oxford OX1 3RH, United Kingdom}
\author{H.~Mendez}
\affiliation{University of Puerto Rico, Mayaguez, Puerto Rico 00681}
\author{J.~Y.~Ge}
\author{D.~H.~Miller}
\author{I.~P.~J.~Shipsey}
\author{B.~Xin}
\affiliation{Purdue University, West Lafayette, Indiana 47907, USA}
\author{G.~S.~Adams}
\author{D.~Hu}
\author{B.~Moziak}
\author{J.~Napolitano}
\affiliation{Rensselaer Polytechnic Institute, Troy, New York 12180, USA}
\author{K.~M.~Ecklund}
\affiliation{Rice University, Houston, Texas 77005, USA}
\author{Q.~He}
\author{J.~Insler}
\author{H.~Muramatsu}
\author{C.~S.~Park}
\author{E.~H.~Thorndike}
\author{F.~Yang}
\affiliation{University of Rochester, Rochester, New York 14627, USA}
\author{M.~Artuso}
\author{S.~Blusk}
\author{S.~Khalil}
\author{R.~Mountain}
\author{K.~Randrianarivony}
\author{T.~Skwarnicki}
\author{S.~Stone}
\author{J.~C.~Wang}
\author{L.~M.~Zhang}
\affiliation{Syracuse University, Syracuse, New York 13244, USA}
\author{G.~Bonvicini}
\author{D.~Cinabro}
\author{M.~Dubrovin}
\author{A.~Lincoln}
\author{M.~J.~Smith}
\author{P.~Zhou}
\author{J.~Zhu}
\author{}
\affiliation{Wayne State University, Detroit, Michigan 48202, USA}
\author{P.~Naik}
\author{J.~Rademacker}
\affiliation{University of Bristol, Bristol BS8 1TL, United Kingdom}
\author{D.~M.~Asner}
\author{K.~W.~Edwards}
\author{J.~Reed}
\author{A.~N.~Robichaud}
\author{G.~Tatishvili}
\author{E.~J.~White}
\affiliation{Carleton University, Ottawa, Ontario, Canada K1S 5B6}
\author{R.~A.~Briere}
\author{H.~Vogel}
\affiliation{Carnegie Mellon University, Pittsburgh, Pennsylvania 15213, USA}
\author{P.~U.~E.~Onyisi}
\author{J.~L.~Rosner}
\affiliation{Enrico Fermi Institute, University of
Chicago, Chicago, Illinois 60637, USA}
\author{J.~P.~Alexander}
\author{D.~G.~Cassel}
\author{R.~Ehrlich}
\author{L.~Fields}
\author{L.~Gibbons}
\author{R.~Gray}
\author{S.~W.~Gray}
\author{D.~L.~Hartill}
\author{B.~K.~Heltsley}
\author{D.~Hertz}
\author{J.~M.~Hunt}
\author{J.~Kandaswamy}
\author{D.~L.~Kreinick}
\author{V.~E.~Kuznetsov}
\author{J.~Ledoux}
\author{H.~Mahlke-Kr\"uger}
\author{J.~R.~Patterson}
\author{D.~Peterson}
\author{D.~Riley}
\author{A.~Ryd}
\author{A.~J.~Sadoff}
\author{X.~Shi}
\author{S.~Stroiney}
\author{W.~M.~Sun}
\author{T.~Wilksen}
\affiliation{Cornell University, Ithaca, New York 14853, USA}
\author{J.~Yelton}
\affiliation{University of Florida, Gainesville, Florida 32611, USA}
\author{P.~Rubin}
\affiliation{George Mason University, Fairfax, Virginia 22030, USA}
\author{N.~Lowrey}
\author{S.~Mehrabyan}
\author{M.~Selen}
\author{J.~Wiss}
\affiliation{University of Illinois, Urbana-Champaign, Illinois 61801, USA}
\author{M.~Kornicer}
\author{R.~E.~Mitchell}
\author{M.~R.~Shepherd}
\author{C.~Tarbert}
\affiliation{Indiana University, Bloomington, Indiana 47405, USA }
\author{D.~Besson}
\affiliation{University of Kansas, Lawrence, Kansas 66045, USA}
\author{T.~K.~Pedlar}
\author{J.~Xavier}
\affiliation{Luther College, Decorah, Iowa 52101, USA}
\author{D.~Cronin-Hennessy}
\author{K.~Y.~Gao}
\author{J.~Hietala}
\author{T.~Klein}
\author{R.~Poling}
\author{P.~Zweber}
\affiliation{University of Minnesota, Minneapolis, Minnesota 55455, USA}
\collaboration{CLEO Collaboration}
\noaffiliation

\date{April 15, 2009}

\begin{abstract} 
We study the inclusive decays of $D^{+}_{s}$ mesons, using data
collected near the $D^{\ast +}_s D^-_s$ peak production energy
$E_{\text{cm}} = 4170$~MeV by the CLEO-c detector. We report the
inclusive yields of $D^{+}_{s}$ decays to
$K^+ X$,
$K^- X$,
$K^{0}_{S} X$, 
$\pi^+ X$, 
$\pi^- X$, 
$\pi^0 X$,
$\eta X$,
$\eta' X$, 
$\phi X$,
$\omega X$ 
and 
$f_0(980) X$, and also decays into pairs of kaons,
$D^+_s \rightarrow K\bar{K} X$. Using these measurements, we obtain an
overview of $D^{+}_{s}$ decays. 
\end{abstract}

\maketitle

The $D^{+}_{s}$ meson, consisting of a $c$ and $\bar{s}$ quark, is the
least extensively studied of the ground state charmed mesons. Here we
present measurements of many inclusive yields from $D^{+}_{s}$ decay,
thereby obtaining an overview of $D^{+}_{s}$ decays. 

Studies of inclusive branching fractions provide strong constraints on
Monte Carlo simulation. On completion of the measurements described
here, we retuned our Monte Carlo decay table. The comparisons of Monte
Carlo and data yields and spectra given below are {\it after} this
retuning.

In addition to providing an improved Monte Carlo decay table, our
results allow some comparisons with expectations.

Data for this analysis were taken at the Cornell Electron Storage 
Ring (CESR) using the CLEO-c general-purpose solenoidal detector,
which is described in detail
elsewhere~\cite{Briere:2001rn,Kubota:1991ww,cleoiiidr,cleorich}. 
The charged
particle tracking system covers a solid angle of 93\% of $4 \pi$
and consists of a small-radius, six-layer, low-mass, stereo wire
drift chamber, concentric with, and surrounded by, a 47-layer
cylindrical central drift chamber. The chambers operate in a 1.0 T
magnetic field and achieve a momentum resolution of $\sim$0.6\%
at $p=$1~GeV/$c$.  
Photons are detected in an electromagnetic calorimeter consisting of
7800 cesium iodide crystals and covering 95\% of $4 \pi$, which
achieves a photon energy resolution of 2.2\% at $E_\gamma=$1~GeV and
6\% at 100~MeV.
We utilize two particle identification (PID) devices to separate
charged kaons from pions: the central drift chamber, which provides
measurements of ionization energy loss ($dE/dx$), and, surrounding
this drift chamber, a cylindrical ring-imaging Cherenkov (RICH)
detector, whose active solid angle is 80\% of $4 \pi$. The combined
PID system has a pion or kaon efficiency $>85\%$ and a probability of
pions faking kaons (or vice versa) $<5\%$~\cite{CLEO:sys}. 
The detector response is modeled with a detailed
GEANT-based~\cite{geant} Monte Carlo (MC) simulation, with initial
particle trajectories generated by EvtGen~\cite{evtgen} and
final state radiation produced by PHOTOS~\cite{photos}.
The initial-state radiation is modeled using cross sections for
$D^{\ast \pm}_s D^\mp_s$ production at lower energies obtained from
the CLEO-c energy scan~\cite{Poling:2006da} near the CM energy where
we collect the sample. 

We use 586 $\mathrm{pb}^{-1}$ of data produced in $e^+ e^-$ collisions 
at CESR near the center-of-mass energy $\sqrt{s}=4170$~MeV. Here the
cross-section for the channel of interest, $D^{\ast +}_s D^-_s$ or
$D^+_s D^{\ast -}_s$, is $\sim$1 nb~\cite{Poling:2006da}. We select
events in which the $D^{\ast}_s$ decays to $D_s \gamma$ (94\%
branching fraction~\cite{PDGValue}). Other charm production totals
$\sim$7 nb~\cite{Poling:2006da}, and the underlying light-quark
``continuum'' is about 12 nb.

Here we employ a double-tagging technique.
Single-tag (ST) events are selected by fully reconstructing a
$D^{-}_{s}$, which we call a tag, in one of the following three
two-body hadronic decay modes: $D_s^- \to K^0_S K^-$, 
$D^-_s \to \phi \pi^-$ and $D^-_s \to K^{\ast 0} K^-$. (Mention of a
specific mode implies the use of the charge conjugate mode as well
throughout this paper.) Details on the tagging selection procedure
are given in Ref.~\cite{CLEO:taunu}.
The tagged $D^-_s$ candidate can be either the primary $D^-_s$ or the
secondary $D^-_s$ from the decay $D^{\ast -}_{s} \to \gamma D^-_s$.
We require the following intermediate states to satisfy these mass
windows around the nominal mass~\cite{PDGValue}: $K^0_S \to \pi^+
\pi^-$ ($\pm 12$ MeV), $\phi \to K^+ K^-$ ($\pm 10$ MeV) and
$K^{\ast 0} \to K^+ \pi^-$ ($\pm 75$ MeV). All charged particles
utilized in tags must
have momenta above 100~MeV/$c$ to eliminate the soft pions from
$D^\ast \bar{D}^\ast$ decays (through $D^\ast \to \pi D$).

We use the reconstructed invariant mass of the $D_s$ candidate,
$M(D_s)$, and the mass recoiling against the $D_s$ candidate, 
$M_\text{recoil}(D_s)
\equiv \sqrt{ (E_{0} - E_{D_s} )^2 - (\mathbf{p}_{0}-\mathbf{p}_{D_s})^2 }
$, as our primary kinematic variables to select a $D_s$
candidate. Here $(E_{0},\mathbf{p}_{0})$ is the net four-momentum of
the $e^+e^-$ beams, taking the finite beam crossing angle into account,
$\mathbf{p}_{D_s}$ is the momentum of the $D_s$ candidate, 
$E_{D_s} = \sqrt{m^2_{D_s} + \mathbf{p}^2_{D_s}}$,
and $m_{D_s}$ is the known $D_s$ mass~\cite{PDGValue}.
We require the recoil mass to be within $55$ MeV of the $D^\ast_s$
mass~\cite{PDGValue}. This loose window allows both primary and
secondary $D_s$ tags to be selected. We also require a photon
consistent with coming from $D^{\ast}_s \rightarrow \gamma D_s$
decay, by looking at the mass recoiling against the $D_s$ candidate
plus $\gamma$ system, $M_\text{recoil}(D_s \gamma)
\equiv \sqrt{ (E_{0} - E_{D_s} - E_\gamma)^2 - (\mathbf{p}_{0}-\mathbf{p}_{D_s}-\mathbf{p}_\gamma)^2 }$.
For correct combinations, this recoil mass peaks at $m_{D_s}$,
regardless of whether the candidate is due to a primary or a secondary
$D_s$. We require 
$| M_\text{recoil}(D_s \gamma) - m_{D_{s}} | < 30~\text{MeV}$.

\begin{figure*}
  \centering{
    \includegraphics*[width=0.95\textwidth]{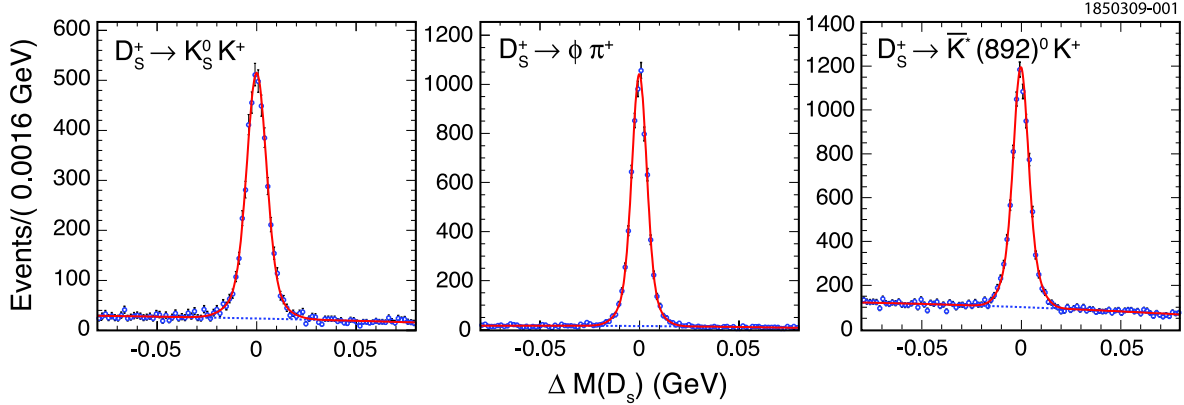} 
  }
  \caption{The mass difference $\Delta M (D_s) \equiv M(D_s) - m_{D_s}$
    distributions in each tag mode.
    We fit the $\Delta M(D_s)$ distribution (open circle)
    to the sum (solid curve)
    of signal (double-Gaussian)
    plus background (second degree polynomial, dashed curve)
    functions. }
  \label{fig:tag}
\end{figure*}

The invariant mass distributions of $D_{s}$ tag candidates for each
tag mode are shown Fig.~\ref{fig:tag}. We use the ST invariant mass
sidebands to estimate the background in our signal yields from
combinatorial background under the ST mass peaks. The signal region is
$|\Delta M(D_s)| < 20$~MeV, while the sideband region is 
$35$~MeV $< |\Delta M(D_s)| < 55$~MeV, where 
$\Delta M(D_s) \equiv M(D_s) - m_{D_s}$ is the difference between the
tag mass and the nominal mass. To find the sideband scaling
factor, the $\Delta M(D_s)$ distributions are fit to the sum of
double-Gaussian signal plus second-degree polynomial background
functions. We have 18586~$\pm$~163 ST events that we use for further
analysis. 

In each event where a tag is identified, we search for our signal
inclusive modes recoiling against the tag. Charged tracks utilized in
signal candidates are required to satisfy criteria based on the track
fit quality, have momenta above 50~MeV/$c$, and angles with respect to
the beam line, $\theta$, satisfying $|\cos\theta|<0.80$. They must
also be consistent with coming from the interaction point in three
dimensions. Pion and kaon candidates are required to have $dE/dx$
measurements within three standard deviations ($3\sigma$) of the
expected value. For tracks with momenta greater than 700~MeV/$c$, RICH
information, if available, is combined with $dE/dx$. Candidate
positrons (and electrons), selected with criteria described in
Ref.~\cite{CLEO:elec}, are required to have momenta of at least
200~MeV/$c$. 

For $D^+_s \rightarrow K^+ X$,
$D^+_s \rightarrow K^- X$,
$D^+_s \rightarrow \pi^+ X$, and
$D^+_s \rightarrow \pi^- X$ modes, we count the numbers of charged
kaons and pions recoiling against the tag where the tags are selected
from both $M(D_s)$ signal
and sideband regions. Thus the combinatoric background is subtracted
by using $M(D_s)$ sideband events. The particle misidentification
backgrounds among $e$, $\pi$ and $K$ are estimated by using the
momentum-dependent particle misidentification rates determined from
Monte Carlo and the $e$, $\pi$ and $K$ yields. Our identification can
not distinguish between muons and pions. So, we assume the muon yield
equals the electron yield, and subtract accordingly.
For $D^+_s \rightarrow \pi^+ X$
and $D^+_s \rightarrow \pi^- X$ modes, we treat $\pi^{\pm}$ from
$K^{0}_{S}$ decay as a background and subtract it based on $K^{0}_{S}$
yields. The momentum-dependent (50 MeV bins) efficiencies for track
finding, track selection criteria, and particle identification are
obtained from Monte Carlo simulation. 

The $K^{0}_{S}$ candidates are reconstructed in $K^{0}_{S} \rightarrow
\pi^+ \pi^-$ decay. The two pions have no PID requirements, and a
vertex fit is done to allow for the $K^{0}_{S}$ flight distance. 
We identify $\pi^0$ candidates via $\pi^0 \rightarrow \gamma \gamma$,
detecting the photons in the CsI calorimeter. We require that the
calorimeter clusters have a measured energy above 30~MeV, have a
lateral distribution consistent with that from photons, and not be
matched to any charged track. The $K^{0}_{S}$ (or $\pi^0$) yield is
extracted by defining a signal region and sideband regions in the
invariant mass distribution of the pion (or photon) pair. The
sideband scaling factor is obtained from Monte Carlo, thus allowing
for a non-linear background shape. We treat
$\pi^{0}$ from $K^{0}_{S}$ decay as a background for the decay 
$D^+_s \rightarrow \pi^0 X$ and subtract it based on $K^{0}_{S}$
yields.

\begin{figure}
  \centering{
    \includegraphics*[width=0.45\textwidth]{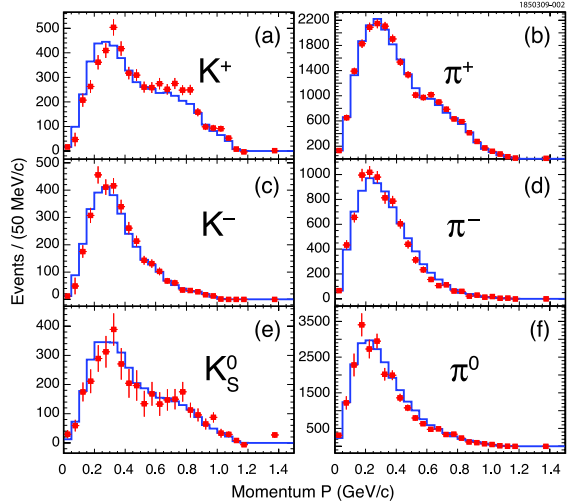}}
  \caption{Charged and neutral kaon and pion momentum spectra after
    background subtractions and efficiency corrections: 
    (a) $D^{+}_{s} \rightarrow K^+ X$,
    (b) $D^{+}_{s} \rightarrow \pi^+ X$,
    (c) $D^{+}_{s} \rightarrow K^- X$,
    (d) $D^{+}_{s} \rightarrow \pi^- X$,
    (e) $D^{+}_{s} \rightarrow K^{0}_{S} X$,
    (f) $D^{+}_{s} \rightarrow \pi^0 X$.
    The points are obtained from data and solid line indicates the
    Monte Carlo after tuning. Good agreement between data and tuned
    Monte Carlo is found. Monte Carlo is normalized to data based on
    tag yield.} 
  \label{fig:pcharg}
\end{figure}

The momentum spectra after all background subtractions and efficiency
corrections are shown in Fig.~\ref{fig:pcharg} 

\begin{figure}
  \centering{
    \includegraphics*[width=0.45\textwidth]{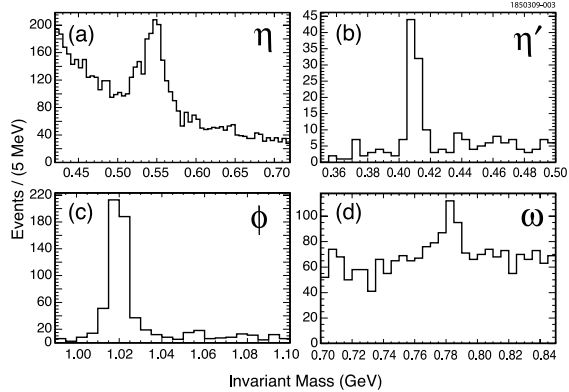} }
  \caption{Invariant mass distributions: 
    (a) $D^{+}_{s} \rightarrow \eta X$,
    (b) $D^{+}_{s} \rightarrow \eta' X$,
    (c) $D^{+}_{s} \rightarrow \phi X$,
    (d) $D^{+}_{s} \rightarrow \omega X$.
  }
  \label{fig:allmass}
\end{figure}

For the $\eta$ we use the $\gamma\gamma$ final state, which has a
large branching fraction in $\eta$ decays. To better handle the mild
dependence of efficiency on $\eta$ momentum, we separate the $\eta$
sample into two momentum ranges to measure the inclusive yields, one
below 300~MeV/$c$ and the other above. The $\eta$ signal
and background yields are determined by fits to a Crystal Ball
function~\cite{CBFunc}, to account for the peak and the low mass tail,
and background polynomial. We reconstruct $\eta'$ candidates in the
the decay mode $\eta' \rightarrow \pi^+ \pi^- \eta$ with the $\eta$ 
subsequently decaying into $\gamma\gamma$. Candidates for $\eta'$ are
selected by combining $\eta$ candidates within 3 r.m.s. widths of the
nominal $\eta$ mass, with a pair of $\pi^+\pi^-$. The mass difference
between $\eta\pi^+\pi^-$ and $\eta$ is then examined and fit to a
Gaussian signal function and a background polynomial to extract the
$\eta'$ yields. The $\phi$ candidates are reconstructed in $\phi
\rightarrow K^+K^-$ decay. We break the $\phi$ sample into several
momentum regions (200~MeV/$c$ bins) since the $\phi$ efficiency
changes substantially with momentum. In each momentum region, the
signals are fit with a sum of two Gausssian shapes and the background
is fit to a polynomial. We reconstruct $\omega$ candidates in $\omega
\rightarrow \pi^+ \pi^- \pi^0$ decay and extract the $\omega$ signal
yields from the $\pi^+ \pi^- \pi^0$ invariant mass distribution. The
invariant mass distributions of $\eta$, $\eta'$, $\phi$, and $\omega$
candidates, summed over all momenta, are shown in
Fig.~\ref{fig:allmass}.

We form
$f_0(980)$ candidates using $\pi^+ \pi^-$ pairs, 
$f_0(980) \rightarrow \pi^+ \pi^-$. The pions are subject
to the standard pion PID requirements. We find no significant evidence
for the decay $D^+_s \rightarrow f_0(980) X$. We fit the invariant
mass distribution of $\pi^+ \pi^-$ pairs to a Gaussian signal function
plus a second-degree polynomial background function and we obtain a
yield of 30 $\pm$ 47. The 90\% confidence level upper limit is 
$\mathcal{B}(D^{+}_{s} \rightarrow f_0(980) X)\mathcal{B}(f_0(980) \rightarrow
\pi^+ \pi^-) < 1.1\%$ (statistical uncertainty only). Systematic
errors are 6.8\% for the efficiency estimation, 5.6\% for the signal
and background shape parameters, and other smaller errors, leading to
a combined relative systematic error of 8.8\%. We conservatively
increase the upper limit by 1.28 times the combined systematic errors,
giving a upper limit, including systematic errors, of
$\mathcal{B}(D^{+}_{s} \rightarrow f_0(980) X)\mathcal{B}(f_0(980) \rightarrow
\pi^+ \pi^-) <1.3\%$. 

We also measure the inclusive yields of $D^+_s$ mesons
into two kaons. After a tag is identified, we search for the best kaon
pair, based on particle identification likelihood or $K^{0}_{S}$ mass,
per mode recoiling against the tag. The kaon pair modes can be
any of $K^{0}_{S}K^{0}_{S}$, $K^{0}_{S}K^+$, $K^{0}_{S}K^-$, $K^+K^-$,
$K^+K^+$ or $K^-K^-$. For
$D^+_s \rightarrow K^{0}_{S}K^{+} X$ and 
$D^+_s \rightarrow K^{0}_{S}K^{-} X$, we apply the sideband
subtraction on $K^{0}_{S}$ candidate invariant mass distribution to
remove the nonresonant decay background and get the signal yields.
The $D^+_s \rightarrow K^{0}_{S}K^{0}_{S} X$
signal yield is extracted by defining a signal region on the scatter
plot for the two $K^{0}_{S}$ candidate invariant masses. In order to
account for $D^+_s \rightarrow K^{0}_{S} \pi^+ \pi^- X$ and 
$D^+_s \rightarrow \pi^+ \pi^- \pi^+ \pi^- X$ 
entering into the signal region of 
$D^+_s \rightarrow K^{0}_{S}K^{0}_{S} X$, we perform a background
subtraction which has two components.
For all two charged kaons modes, we count the event numbers
where at lease two charged kaons are found recoiling against the tag.
In order to subtract the combinatoric background, we repeat the same
procedure for each mode where the tags are selected from $M(D_s)$
sidebands. The other possible backgrounds from generic $D_s$ decay are
studied using Monte Carlo and found to be negligible.

The double-tagging technique allows us to measure the inclusive
yields for the decay $D^+_s \rightarrow K^{0}_{L} X$
without directly detecting the $K^{0}_{L}$. Instead, we reconstruct
all particles in the event except the single $K^{0}_{L}$ and infer the
presence of a $K^{0}_{L}$ from the missing four-momentum. Our signal
is a peak in the missing mass squared distribution at the $K^{0}_{L}$
mass squared. Similar missing-mass-squared techniques are used for
$D^+_s \rightarrow K^{0}_{L} K^{0}_{S} X$,
$D^+_s \rightarrow K^{0}_{L} K^+ X$ and 
$D^+_s \rightarrow K^{0}_{L} K^- X$ modes by requiring there must be
a $K^{0}_{S}$, $K^+$ or $K^-$ recoiling against the tag. Note that if the
$D_s$ decay contains two or more $K^{0}_{L}$'s, we do not find any
$K^{0}_{L}$. Due to the low statistics and large systematic
uncertainties, we quote the inclusive $K^{0}_{L}$ results only as a
check for $K^{0}_{S}$.

\begin{table*}[tb]
  \begin{center}
    \caption{\label{tab:all} $D_s$ inclusive yield
      results. Uncertainties are statistical and systematic,
      respectively. The inclusive $K^{0}_{L}$ results are only used as
      a check for $K^{0}_{S}$. The $D^+_s \rightarrow K^{0}_{L} X$
      yield requires a correction before comparing with the $D^+_s
      \rightarrow K^{0}_{S} X$ yield, as explained in the text.
      PDG~\cite{PDGValue} averages are shown
      in the last column, when available.}
    \begin{tabular}{ l rclcl l rcl rcl }
      \hline \hline
      Mode
      & \multicolumn{5}{c}{Yield(\%)}
      & ~~~~$K^{0}_{L}$ Mode
      & \multicolumn{3}{c}{Yield(\%)}
      & \multicolumn{3}{r}{$\mathcal{B}$(PDG)(\%)}  \\ \hline

      $D^+_s \rightarrow $$\pi^+ X$ & 119.3 & $\pm$ & 1.2 & $\pm$ & 0.7 & & & & & & &   \\
      $D^+_s \rightarrow $$\pi^- X$ & 43.2 & $\pm$ & 0.9 & $\pm$ & 0.3 & & & & & & &   \\
      $D^+_s \rightarrow $$\pi^0 X$ & 123.4 & $\pm$ & 3.8 & $\pm$ & 5.3 & & & & & & &   \\
      $D^+_s \rightarrow $$K^+ X$ & 28.9 & $\pm$ & 0.6 & $\pm$ & 0.3 & & & & & 20 & $^{+}_{-}$ & $^{18}_{14}$   \\
      $D^+_s \rightarrow $$K^- X$ & 18.7 & $\pm$ & 0.5 & $\pm$ & 0.2 & & & & & 13 & $^{+}_{-}$ & $^{14}_{12}$   \\
      $D^+_s \rightarrow $$\eta X$ & 29.9 & $\pm$ & 2.2 & $\pm$ & 1.7 & & & & & & &   \\
      $D^+_s \rightarrow $$\eta' X$ & 11.7 & $\pm$ & 1.7 & $\pm$ & 0.7 & & & & & & &   \\
      $D^+_s \rightarrow $$\phi X$ & 15.7 & $\pm$ & 0.8 & $\pm$ & 0.6 & & & & & & &   \\
      $D^+_s \rightarrow $$\omega X$ & 6.1 & $\pm$ & 1.4 & $\pm$ & 0.3 & & & & & & &   \\
      $D^+_s \rightarrow $$f_0(980) X, f_0(980) \rightarrow \pi^+\pi^-$ &  \multicolumn{5}{c}{ $<$ 1.3\% (90\%  CL) } & & & & & & &   \\
      $D^+_s \rightarrow $$K^{0}_{S} X$ &  19.0 & $\pm$ & 1.0 & $\pm$ & 0.4  & $D^+_s \rightarrow $$K^{0}_{L} X$  &  15.6 & $\pm$ & 2.0 &  20 & $\pm$ & 14   \\
      $D^+_s \rightarrow $$K^{0}_{S} K^{0}_{S} X$ & 1.7 & $\pm$ & 0.3 & $\pm$ & 0.1 & $D^+_s \rightarrow $$K^{0}_{L} K^{0}_{S} X$ &  5.0 & $\pm$ & 1.0 & & &   \\
      $D^+_s \rightarrow $$K^{0}_{S} K^+ X$ & 5.8 & $\pm$ & 0.5 & $\pm$ & 0.1 & $D^+_s \rightarrow $$K^{0}_{L} K^+ X$ &  5.2 & $\pm$ & 0.7 & & & \\
      $D^+_s \rightarrow $$K^{0}_{S} K^- X$ & 1.9 & $\pm$ & 0.4 & $\pm$ & 0.1 & $D^+_s \rightarrow $$K^{0}_{L} K^- X$ &  1.9 & $\pm$ & 0.3 & & & \\
      $D^+_s \rightarrow $$K^+ K^- X$ & 15.8 & $\pm$ & 0.6 & $\pm$ & 0.3 & & & & & & &   \\
      $D^+_s \rightarrow $$K^+ K^+ X$ &  \multicolumn{5}{c}{ $<$ 0.26\% (90\%  CL) } & & & & & & &   \\
      $D^+_s \rightarrow $$K^- K^- X$ &  \multicolumn{5}{c}{ $<$ 0.06\% (90\%  CL) } & & & & & & &   \\
      \hline \hline
    \end{tabular}
  \end{center}
\end{table*}      

The inclusive yields are listed in
Table~\ref{tab:all}. For the $K^{0}_{S}$ modes, the corresponding
$K^{0}_{L}$ modes are listed as a comparison. The value of the decay
$D^+_s \rightarrow K^{0}_{L} X$ is only for $D^+_s$ 
decaying into a single $K^{0}_{L}$. So one should not directly
compare the values of $D^+_s \rightarrow K^{0}_{S} X$ and 
$D^+_s \rightarrow K^{0}_{L} X$ in Table~\ref{tab:all}. One can
correct the single $K^{0}_{L}$ inclusive yield by adding two times
the inclusive yield of 
$D^+_s \rightarrow K^{0}_{L} K^{0}_{L} X$ (assuming
$\mathcal{B}(D^+_s \rightarrow K^{0}_{L} K^{0}_{L} X) = 
\mathcal{B}(D^+_s \rightarrow K^{0}_{S} K^{0}_{S} X)$). All the
$K^{0}_{L}$ modes are consistent with $K^{0}_{S}$ modes. In the last
column of Table~\ref{tab:all}, we show PDG~\cite{PDGValue} averages,
when available.

We have considered several sources of systematic uncertainty.
The uncertainty associated with the efficiency for
finding a track is 0.3\%; an additional 0.6\% systematic
uncertainty for each kaon track is added~\cite{CLEO:sys}.
The relative systematic uncertainties for
$\pi^0$ and $K^0_S$ efficiencies are 4.2\% and  
1.8\%, respectively. Uncertainties in the charged pion and kaon
identification efficiencies are 0.3\% per pion and 0.3\% per kaon~\cite{CLEO:sys}.
All efficiencies from Monte Carlo have been
corrected to include several known small differences between
data and Monte Carlo simulation.

\begin{figure}
  \centering{
    \includegraphics*[width=0.45\textwidth]{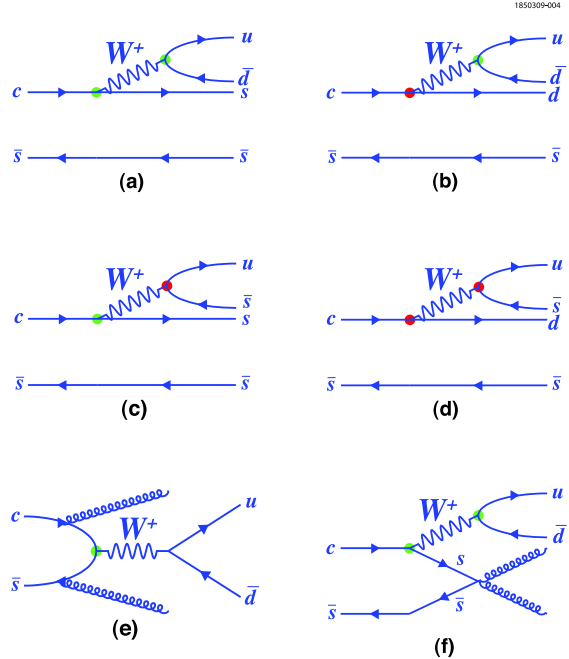} }
  \caption{The typical Feynman diagrams of $D^{+}_{s}$ decays: 
    (a) Cabibbo-favored decay, 
    (b) single-Cabibbo-suppressed decay,
    (c) single-Cabibbo-suppressed decay,
    (d) double-Cabibbo-suppressed decay,
    (e) short-range annihilation decay,
    (f) long-range annihilation decay. }
  \label{fig:dia}
\end{figure}

The quark-level diagrams contributing to $D^{+}_{s}$ decay are shown in
Fig.~\ref{fig:dia}. We classify ``quark-level final states'' as
$s\bar{s}$ (as would come from Fig.~\ref{fig:dia}(a)), $\bar{s}$
(Fig.~\ref{fig:dia}(b)), $s\bar{s}\bar{s}$ (Fig.~\ref{fig:dia}(c)),
$\bar{s}\bar{s}$ (Fig.~\ref{fig:dia}(d)), and ``no strange quarks''
(Fig.~\ref{fig:dia}(e) and Fig.~\ref{fig:dia}(f)). The $s\bar{s}$ final
state is Cabibbo-favored. The $\bar{s}$ and $s\bar{s}\bar{s}$ final
states are singly-Cabibbo-suppressed, the $\bar{s}\bar{s}$ final
state is doubly-Cabibbo-suppressed, and the ``no strange quarks''
final state arises from short-range (Fig.~\ref{fig:dia}(e)) and
long-range (Fig.~\ref{fig:dia}(f)) annihilation diagrams (While
Fig.~\ref{fig:dia}(f) shows the $s\bar{s}$ annihilating into gluons,
here we also include its rescattering into $u\bar{u}$ or $d\bar{d}$).

The $s\bar{s}$ final state can hadronize as $K\bar{K} X$, but
also as $\eta X$, $\eta' X$, or $\phi X$. The $\bar{s}$ final state
will hadronize as $K X$. The $s\bar{s}\bar{s}$ final state in
principle can hadronize as $KK\bar{K} X$, but there will be limited
phase space for this, so $K\eta X$, $K\eta' X$, $K\phi X$ are probably
more likely. The $\bar{s}\bar{s}$ final state will hadronize as 
$KK X$, but being doubly-Cabibbo-suppressed, can probably be ignored.

We have performed a global fit to our measurements. For this, we have
branching fractions $\mathcal{B}(XX)$. In particular, for $s\bar{s}$
quark-level final states, we write 
$\mathcal{B}(D_s \rightarrow s\bar{s}) \equiv \mathcal{B}(s\bar{s})$, 
$\mathcal{B}(D_s \rightarrow s\bar{s} \rightarrow \eta X) \equiv \mathcal{B}(\eta)$, 
$\mathcal{B}(D_s \rightarrow s\bar{s} \rightarrow \eta' X) \equiv \mathcal{B}(\eta')$, 
$\mathcal{B}(D_s \rightarrow s\bar{s} \rightarrow \phi X) \equiv \mathcal{B}(\phi)$, 
and 
$\mathcal{B}(D_s \rightarrow s\bar{s} \rightarrow K\bar{K} X) \equiv \mathcal{B}(K\bar{K})$.
Thus $\mathcal{B}(s\bar{s}) = \mathcal{B}(\eta) +
\mathcal{B}(\eta') + \mathcal{B}(\phi) + \mathcal{B}(K\bar{K})$. 
Note that $\mathcal{B}(D_s \rightarrow s\bar{s} \rightarrow \eta X)$
is the branching fraction for {\it primary} production of $\eta$ 
({\it not} from $\eta'$ decay), from the quark-level state
$s\bar{s}$. The free parameters in our fit are $\mathcal{B}(\eta)$,
$\mathcal{B}(\eta')$, $\mathcal{B}(\phi)$, and
$\mathcal{B}(K\bar{K})$, which we adjust to obtain the best fit. 

For the $\bar{s}$ quark-level final state, we note that 
$\mathcal{B}(D_s \rightarrow \bar{s}) \equiv \mathcal{B}(\bar{s})
\approx |V_{\text{cd}}/V_{\text{cs}}|^{2} \times
\mathcal{B}(s\bar{s})$. Thus, we do not adjust $\mathcal{B}(\bar{s})$
in the fit, but write $\mathcal{B}(\bar{s}) = C_1 \times
|V_{\text{cd}}/V_{\text{cs}}|^{2} \times \mathcal{B}(s\bar{s})$, where
$C_1$ is a phase space correction factor, probably a bit larger than
1.0. We take $C_1$ to be $1.25 \pm 0.25$.

We break the $s\bar{s}\bar{s}$ quark-level final state into 4 separate pieces, as
we have done with the $s\bar{s}$ final state. Thus
$\mathcal{B}(D_s \rightarrow s\bar{s}\bar{s}) \equiv \mathcal{B}(s\bar{s}\bar{s})$
is made up of 
$\mathcal{B}(D_s \rightarrow s\bar{s}\bar{s} \rightarrow \eta\bar{s} X) \equiv \mathcal{B}(\eta\bar{s})$,
$\mathcal{B}(D_s \rightarrow s\bar{s}\bar{s} \rightarrow \eta'\bar{s} X) \equiv \mathcal{B}(\eta'\bar{s})$,
$\mathcal{B}(D_s \rightarrow s\bar{s}\bar{s} \rightarrow \phi\bar{s} X) \equiv \mathcal{B}(\phi\bar{s})$,
and 
$\mathcal{B}(D_s \rightarrow s\bar{s}\bar{s} \rightarrow K\bar{K}\bar{s} X) \equiv \mathcal{B}(K\bar{K}\bar{s})$. Thus
$\mathcal{B}(s\bar{s}\bar{s}) = \mathcal{B}(\eta\bar{s}) +
\mathcal{B}(\eta'\bar{s}) + \mathcal{B}(\phi\bar{s}) +
\mathcal{B}(K\bar{K}\bar{s})$. We note that
$\mathcal{B}(s\bar{s}\bar{s}) \approx |V_{\text{us}}/V_{\text{ud}}|^{2}
\times \mathcal{B}(s\bar{s})$. So again, we do not adjust any of the
pieces making up $\mathcal{B}(s\bar{s}\bar{s})$, but rather write
\begin{equation}
  \mathcal{B}(\eta\bar{s}) = C_2  \times
  |V_{\text{us}}/V_{\text{ud}}|^{2} \times  
  \mathcal{B}(\eta)
\end{equation}
\begin{equation}
  \mathcal{B}(\eta'\bar{s}) = C_2  \times
  |V_{\text{us}}/V_{\text{ud}}|^{2} \times  
  \mathcal{B}(\eta')
\end{equation}
\begin{equation}
  \mathcal{B}(\phi\bar{s}) = C_2  \times
  |V_{\text{us}}/V_{\text{ud}}|^{2} \times  
  \mathcal{B}(\phi)
\end{equation}
\begin{equation}
  \mathcal{B}(K\bar{K}\bar{s}) = C_2  \times
  |V_{\text{us}}/V_{\text{ud}}|^{2} \times  
  \mathcal{B}(K\bar{K})
\end{equation}
The quantity $C_2$, like $C_1$, is a phase space correction factor,
expected to be smaller than 1.0. We take it to be $0.75 \pm
0.25$. Assuredly the true phase space correction factors would be
different for $\eta$, $\eta'$, $\phi$, and $K\bar{K}$. We neglect this
in our fit, allowing for it as a systematic error.

For the doubly-Cabibbo-suppressed decays, we estimate
$\mathcal{B}(D_s \rightarrow \bar{s}\bar{s}) \equiv 
\mathcal{B}(\bar{s}\bar{s}) = C_3  \times
|(V_{\text{cd}}/V_{\text{cs}})(V_{\text{us}}/V_{\text{ud}})|^2 \times
\mathcal{B}(s\bar{s})$. This term is down a factor of 400 from the
dominant term, and has essentially no effect on our fit. We take 
$C_3 = 1.0 \pm 1.0$.

Finally, there are annihilation diagrams. We write 
$\mathcal{B}(\text{Annihilation}) = 
\mathcal{B}(D^+_s \rightarrow \mu^+  \nu) +
\mathcal{B}(D^+_s \rightarrow \tau^+ \nu) +
\mathcal{B}(D^+_s \rightarrow \text{Other~Annihilation})$. One of our
goals in performing the global fit is to get an estimate of
$\mathcal{B}(D^+_s \rightarrow \text{Other~Annihilation})$. 
In our fit, we use 
$\mathcal{B}(D^+_s \rightarrow \tau^+ \nu)= (5.62  \pm 0.41  \pm 0.16
)\%$~\cite{CLEO:taunu}, and
$\mathcal{B}(D^+_s \rightarrow \mu^+ \nu) = (0.565 \pm 0.045 \pm
0.017)\%$~\cite{CLEO:munu}.

It is possible for a $D_s$ decay to contain more than one of $\eta$,
$\eta'$, $\phi$, $K\bar{K}$, {\it e.g.} $\eta\eta$, $\eta\phi$,
etc. From energy conservation, one of an allowed pair must be
$\eta$. So, we include a yield $\mathcal{B}(\text{extra}~\eta)$ to
allow for this. We searched for 
$D^{+}_{s} \rightarrow \eta\eta X$, 
$D^{+}_{s} \rightarrow \eta\eta' X$, and 
$D^{+}_{s} \rightarrow \eta\phi X$. We found no clear signals,
obtaining a summed yield of ($6.0 \pm 3.9$)\%. In our global fit, we
take $\mathcal{B}(\text{extra}~\eta)$ to be 6.0\%, and include the
$\pm 3.9\%$ in the systematic error.

Another source of $\eta$ and $\eta'$ is the quark-level decay 
$D_s \rightarrow \bar{s}$ (Fig.~\ref{fig:dia}(b)). Here, the $\eta$ or $\eta'$ will come not
from their $s\bar{s}$ component, but from their $u\bar{u}$ and
$d\bar{d}$ components. At quark level, the decay is 
$D_s \rightarrow u\bar{d}d\bar{s}$, so making $\eta$ or $\eta'$ is
natural. We assume that this diagram gives an $\eta$ a fraction $f_1$
of the time, and an $\eta'$ a fraction $f_2$ of the time, where 
$f_1+f_2 \le 1$. While one can make quark-level predictions of what to
expect for $f_1$ and $f_2$, we take the conservative position of
allowing them the full range, $0 \le f_1+f_2 \le 1$, and take
$f_1=f_2=1/4$, in the middle of the allowed range.

For our global fit, we write
{\large
\begin{widetext}
  \begin{equation}
    \label{eq:chi2}
    \begin{array}{cl}
      \chi^2 =
      &(\frac{ \text{Y}_{\eta} -
        (
        \mathcal{B}(\eta)+\mathcal{B}(\eta\bar{s}) + 
        \mathcal{B}(\eta' \rightarrow \eta X) \times (
        \mathcal{B}(\eta')+\mathcal{B}(\eta'\bar{s})+ f_2 \times \mathcal{B}(\bar{s}))+
        \mathcal{B}(\text{extra}~\eta)+
        f_1 \times \mathcal{B}(\bar{s})
        ) }
      {\delta_{\text{Y}_{\eta}}})^2
      + \\
      &(\frac{ \text{Y}_{\eta'} -
        (
        \mathcal{B}(\eta')+\mathcal{B}(\eta'\bar{s})+
        f_2 \times \mathcal{B}(\bar{s})
        ) }
      {\delta_{\text{Y}_{\eta'}}})^2
      + \\
      &(\frac{ \text{Y}_{\phi} -
        (
        \mathcal{B}(\phi)+\mathcal{B}(\phi\bar{s})
        ) }
      {\delta_{\text{Y}_{\phi}}})^2
      + \\
      &(\frac{ \text{Y}_{KK} -
        (
        \mathcal{B}(K\bar{K})+\mathcal{B}(K\bar{K}\bar{s}) + 
        \mathcal{B}(\phi \rightarrow K\bar{K}) \times (
        \mathcal{B}(\phi)+\mathcal{B}(\phi\bar{s})) + 
        \mathcal{B}(\bar{s}\bar{s})
        ) }
      {\delta_{\text{Y}_{KK}}})^2
      + \\
      &(\frac{ \text{Y}_{K} - 
        (
        2\times(\mathcal{B}(K\bar{K})+\mathcal{B}(K\bar{K}\bar{s})) +
        2\times\mathcal{B}(\phi \rightarrow K\bar{K}) \times (
        \mathcal{B}(\phi)+\mathcal{B}(\phi\bar{s})) +
        \mathcal{B}(s\bar{s}\bar{s}) +
        \mathcal{B}(\bar{s}) +
        2\times\mathcal{B}(\bar{s}\bar{s})
        )}
      {\delta_{\text{Y}_{K}}})^2
    \end{array}
  \end{equation}
\end{widetext}
}

Here $\text{Y}_{i}$ is the central value of a measurement, and 
$\delta_{\text{Y}_{i}}$ is the error on that  measurement. As $\eta'$
decays to $\eta$, and $\phi$ decays to $K\bar{K}$, our $\chi^2$ needs
the branching fractions for those decays, 
$\mathcal{B}(\eta' \rightarrow \eta X)$ and 
$\mathcal{B}(\phi \rightarrow K \bar{K})$. We take these from
PDG~\cite{PDGValue}. Better than words, Eq.~(\ref{eq:chi2}) gives the
meaning of the various $\mathcal{B}(XX)$ parameters. Thus, the
measured yield of $\eta$, $\text{Y}_{\eta}$, has contributions from
primary production of $\eta$ from the $s\bar{s}$ quark state
($\mathcal{B}(\eta)$), primary production of $\eta$ from the
$s\bar{s}\bar{s}$ quark state ($\mathcal{B}(\eta\bar{s})$), primary
production of $\eta$ from the $\bar{s}$ quark state 
($f_1 \times \mathcal{B}(\bar{s})$), production
of $\eta$ from decay of $\eta'$, the $\eta'$ being from the $s\bar{s}$
quark state ($\mathcal{B}(\eta') \times \mathcal{B}(\eta' \rightarrow
\eta X)$), or the $\eta'$ being from the $s\bar{s}\bar{s}$ quark state
($\mathcal{B}(\eta'\bar{s}) \times \mathcal{B}(\eta' \rightarrow \eta X)$), 
or from the $\bar{s}$ quark state 
($f_2 \times \mathcal{B}(\bar{s}) \times \mathcal{B}(\eta' \rightarrow \eta X)$),
and finally of ``extra $\eta$'s'', $\eta$ that accompanies an $\eta$,
$\eta'$, or $\phi$ already
recorded ($\mathcal{B}(\text{extra}~\eta)$). The measured yields for
$\eta'$ and $\phi$, while not as complicated, have some of the same
features. Note that, as described earlier, our measured yield of
di-kaons, $\text{Y}_{KK}$, includes $K\bar{K}$ and $KK$ and
$\bar{K}\bar{K}$ pairs. There is a subtlety in the last line of
Eq.~(\ref{eq:chi2}). The decay $D_s \rightarrow s\bar{s}\bar{s}$ always
makes at least one kaon, and when the decay is 
$D_s \rightarrow K\bar{K}\bar{s}$, i.e.,
$\mathcal{B}(K\bar{K}\bar{s})$, makes 2 more. Line 5, for the kaon
yield, properly handles this.

We minimize $\chi^2$ by varying $\mathcal{B}(\eta)$,
$\mathcal{B}(\eta')$, $\mathcal{B}(\phi)$, and
$\mathcal{B}(K\bar{K})$. All other $\mathcal{B}(XX)$ parameters are
fixed as previously described. Further, we have the unitarity
requirement $\mathcal{B}(s\bar{s}) + \mathcal{B}(s\bar{s}\bar{s}) +
\mathcal{B}(\bar{s}) + \mathcal{B}(\bar{s}\bar{s}) +
\mathcal{B}(\text{Annihilation}) = 1.0$. Our fit gives
$\mathcal{B}(\eta)$, $\mathcal{B}(\eta')$, $\mathcal{B}(\phi)$,
$\mathcal{B}(K\bar{K})$, and hence $\mathcal{B}(s\bar{s})$,
$\mathcal{B}(s\bar{s}\bar{s})$, $\mathcal{B}(\bar{s})$, and
$\mathcal{B}(\bar{s}\bar{s})$. Unitarity then gives
$\mathcal{B}(\text{Other~Annihilation})$. Results are given in
Table~\ref{tab:fitpa}.

\begin{table}
  \begin{center}
    \caption{\label{tab:fitpa}Results from the global fit. The central
      values of parameters are listed in second column. The errors:
      $\delta_1$ is statistical uncertainty, 
      $\delta_2$ is from phase space factor $C_1=1.25\pm0.25$,
      $\delta_3$ is from phase space factor $C_2=0.75\pm0.25$,
      $\delta_4$ is from $f_1+f_2=0.5\pm0.5$, and
      $\delta_5$ is from the
      $\mathcal{B}(\text{extra~\eta})=(6.0\pm3.9)\%$.}
    \begin{tabular}{l c c c c c c}
      \hline \hline
      Parameter&Value(\%)&\multicolumn{5}{c}{Error(\%)} \\
      &&$\delta_1$&$\delta_2$&$\delta_3$&$\delta_4$&$\delta_5$
      \\ \hline
      $\mathcal{B}(D_s \rightarrow s\bar{s} \rightarrow \eta X)$
      & 14.7 & ~2.9~ & ~0.2~ & ~0.2~ & ~1.0~ & ~3.7~\\
      $\mathcal{B}(D_s \rightarrow s\bar{s} \rightarrow \eta' X)$    
      & 10.3 & ~1.7~ & ~0.2~ & ~0.1~ & ~1.0~ & ~0.1~\\
      $\mathcal{B}(D_s \rightarrow s\bar{s} \rightarrow \phi X)$     
      & 15.1 & ~1.0~ & ~0.0~ & ~0.2~ & ~0.0~ & ~0.0~\\
      $\mathcal{B}(D_s \rightarrow s\bar{s} \rightarrow K\bar{K} X)$ 
      & 25.4 & ~1.2~ & ~0.3~ & ~0.6~ & ~0.1~ & ~0.1~\\
      $\mathcal{B}(D_s \rightarrow s\bar{s})$ 
      & 65.6 & ~2.7~ & ~0.7~ & ~1.0~ & ~1.8~ & ~3.5\\ \hline
      $\mathcal{B}(\text{Other~Annihilation})$
      & 21.5 & ~2.8~ & ~0.1~ & ~0.3~ & ~2.0~ & ~3.9~\\
      \hline \hline
    \end{tabular}
  \end{center}
\end{table}      

We have five measurements, and four free parameters. So it would
appear that there is one degree of freedom. However, the single kaon
and di-kaon measurements are highly correlated, so we effectively have
more like four measurements. This is reflected in the $\chi^2$ of the
fit, which is 0.03. We have also made a fit leaving the di-kaon term
out, and a fit leaving the single kaon term out. These fits give
essentially the same result as the nominal fit with both terms
included. 

In interpreting the results in Table~\ref{tab:fitpa}, it should be
recognized that the decay products of the true ``other annihilation''
diagrams will include some $D_s \rightarrow \text{gluons} \rightarrow
s\bar{s}$ events, thus being treated as part of
$\mathcal{B}(s\bar{s})$ rather than ``other annihilation''. Also, the
gluons will make $u\bar{u}$, $d\bar{d}$, which will sometimes make
$\eta$, $\eta'$, again being treated as a contribution to
$\mathcal{B}(s\bar{s})$. Thus $\mathcal{B}(\text{Other~Annihilation})$
should be viewed as a lower bound, $\mathcal{B}(\eta)$,
$\mathcal{B}(\eta')$, $\mathcal{B}(\phi)$, $\mathcal{B}(K\bar{K})$ as
upper bounds, on contributions from the various diagrams in
Fig.~\ref{fig:dia}. On the other hand, an overestimate of
$\mathcal{B}(\text{extra~\eta})$ will give an overestimate of
$\mathcal{B}(\text{Other~Annihilation})$.

We can obtain a conservative lower bound on
$\mathcal{B}(\text{Other~Annihilation})$ by setting $f_1=f_2=0$ and
$\mathcal{B}(\text{extra~\eta})=0$. That gives
$\mathcal{B}(\text{Other~Annihilation})=13.3\pm3.0\%$, i.e., $>9.5\%$ at 90\%
C.L..

We use our measurements of the total kaon yield and the total di-kaon
yield to get a measurement of the singly-Cabibbo-suppressed rate. If
there were no tri-kaon events, then (total kaon yield) minus
$2\times$(total di-kaon yield) would give (single kaon yield) which
would include the $\bar{s}$ final state, and that fraction of the
$s\bar{s}\bar{s}$ final state for which the $s\bar{s}$ component
hadronized as $\eta$, $\eta'$, or $\phi$. Tri-kaon events complicate
the situation. As mentioned earlier, in counting di-kaons, a given
charge pairing ($K^+K^+$, $K^+K^{0}_{S}$, $K^+K^-$ etc.) is counted
once. Thus $K^{0}_{S}K^{0}_{S}K^{0}_{S} X$ is counted as one di-kaon,
while $K^+K^{0}_{S}K^{0}_{S} X$ is counted as two, $K^+K^{0}_{S}K^- X$
as three. For the total kaon yield, a tri-kaon event is counted as 3
kaons, In taking (total kaon yield) minus $2\times$(total di-kaon
yield) as a way of counting singly-Cabibbo-suppressed yield, the
``right'' answer for a tri-kaon event is $+1$, and what we actually
obtain is $+1$, $-1$, and $-3$, for the different tri-kaon events, on
average $-1$ instead of $+1$. Thus, our proposed procedure will
underestimate the singly-Cabibbo-suppressed rate. To the extent that
the tri-kaon rate is small, the underestimate is small. We estimate
and apply a correction.

Our numbers
are: total kaon yield is ($85.6 \pm 2.3$)\%, total di-kaon yield is
($39.9 \pm 1.8$)\%. The errors are {\it highly} correlated. Taking
correlations into consideration, we find
kaon~$-~2\times$di-kaon is ($5.8 \pm 2.2$)\%. Taking
$\mathcal{B}(s\bar{s}\bar{s})/\mathcal{B}(s\bar{s})$ to be $\sim
1/20$, and $\mathcal{B}(s\bar{s}\bar{s} \rightarrow
\text{tri}$-kaon$)/\mathcal{B}(s\bar{s}\bar{s})$ to be
$<\mathcal{B}(K\bar{K})/\mathcal{B}(s\bar{s})=0.39$,
our correction factor for the presence of tri-kaon decays is 
$<(65.6\times\frac{1}{20}\times0.39\times2)\%$.
Thus, the correction factor is $<2.6\%$. Taking it to be $(1.3\pm1.3)\%$,
the measured branching fraction for $D_s \rightarrow$
single-Cabibbo-suppressed is ($7.1 \pm 2.2 \pm 1.3$)\%. The expected
branching fraction is 
$(|V_{\text{us}}/V_{\text{ud}}|^2+|V_{\text{cd}}/V_{\text{cs}}|^2)\times\mathcal{B}(s\bar{s})\approx\frac{1}{10}\times\mathcal{B}(s\bar{s})$.
Taking $\mathcal{B}(s\bar{s})$ from Table~\ref{tab:fitpa}, we see fine
agreement between expectations and measurements.

From our global fit, we can compute the minimum yields of $\pi^+$,
$\pi^-$, and $\pi^0$ for each category. For example, for the
Cabibbo-favored decay $D^+_s \rightarrow s\bar{s} \rightarrow \eta X$, 
with 14.7\% yield, we compute the yields of
$\pi^+$, $\pi^-$, and $\pi^0$ that come from a 14.7\% $\eta$ yield. To
this we add 14.7\% $\pi^+$ yield, since that must be present to
conserve charge. (This is an overestimate, because semileptonic decays
have charge conserved via $e^+$ or $\mu^+$, consequently we perform a
subtraction to allow for that.)
For $D^+_s \rightarrow s\bar{s}\bar{s} \rightarrow \eta \bar{s} X$,
with 0.6\% yield, similarly we compute the yields of $\pi^+$, $\pi^-$,
and $\pi^0$ that come from a 0.6\% $\eta$ yield. Charge conservation
might be achieved by a $\pi^+$, but also by a $K^+$. Lacking any
information on how much comes from $\pi^+$, how much from $K^+$, we
assume half from each. Our global fit gives a single number
$\mathcal{B}(K\bar{K}) = 25.4\%$, for the di-kaon yield. To determine
the $\pi^+$, $\pi^0$, and $\pi^-$ yields, we need yields for the
separate di-kaon combinations, $K^{0}_{S}K^{0}_{S}$, $K^{0}_{S}K^{+}$,
$K^{0}_{S}K^{-}$, etc. For our calculation, we take the measured
di-kaon yields from Table~\ref{tab:all}, and normalize them so their
sum equals $\mathcal{B}(K\bar{K})$. (Where we have only an upper
limit, we use half of it for the ``measurement'').

The results of our computation are given in
Table~\ref{tab:minpi}. There one sees that the yields of $\pi^+$,
$\pi^-$, and $\pi^0$ should be larger than 96.2\%, 20.5\%, and 46.8\%,
respectively. The observed yields are indeed larger than these
numbers. Thus, on average, 1/4 of the $D_s$ decays will contain an
additional $\pi^+\pi^-$ pair, and 3/4 of the $D_s$ decays will contain
an additional $\pi^0$ (or 1/2 contain one additional $\pi^0$, 1/8
contain two additional $\pi^0$'s).

\begin{table*}[tb]      
  \begin{center}
    \caption{\label{tab:minpi}The minimum yields of $\pi^+$, $\pi^-$,
      and $\pi^0$ for each category. We compute the yields of $\pi^+$,
      $\pi^-$, and $\pi^0$ that come from signal particles. In
      addition to that, we add charged pions to conserve charge. 
      Semileptonic decays have charge conserved via $e^+$ or $\mu^+$,
      consequently we perform a subtraction to allow for that.}
    \begin{tabular}{ l  c  c c  c c c  c c c }\hline \hline
      && \multicolumn{2}{c}{Charge} 
      & \multicolumn{3}{c}{Particle}
      & \multicolumn{3}{c}{Total}    \\
      && \multicolumn{2}{c}{Conservation} 
      & \multicolumn{3}{c}{Decay}
      & \multicolumn{3}{c}{Yields}    \\
      Mode & $\mathcal{B}$ (\%)
      &~~ $\pi^+$ ~~&~~ $\pi^-$ ~~
      &~~ $\pi^+$ ~~&~~ $\pi^-$ ~~&~~ $\pi^0$ ~~
      &~~ $\pi^+$ ~~&~~ $\pi^-$ ~~&~~ $\pi^0$ ~~ \\ \hline

      $D^+_s \rightarrow \eta X$ & 14.7 & 14.7 & 0.0 & 4.0 & 4.0 & 17.7 & 18.7 & 4.0 & 17.7 \\
      $D^+_s \rightarrow \eta\bar{s} X$ & 0.6 & 0.3 & 0.0 & 0.2 & 0.2 & 0.7 & 0.4 & 0.2 & 0.7 \\
      $D^+_s \rightarrow \eta' X$ & 10.3 & 10.3 & 0.0 & 9.7 & 9.7 & 12.7 & 20.0 & 9.7 & 12.7 \\
      $D^+_s \rightarrow \eta'\bar{s} X$ & 0.4 & 0.2 & 0.0 & 0.4 & 0.4 & 0.5 & 0.6 & 0.4 & 0.5 \\
      $D^+_s \rightarrow \phi X$ & 15.1 & 15.1 & 0.0 & 2.4 & 2.4 & 2.5 & 17.5 & 2.4 & 2.5 \\
      $D^+_s \rightarrow \phi\bar{s} X$ & 0.6 & 0.3 & 0.0 & 0.1 & 0.1 & 0.1 & 0.4 & 0.1 & 0.1 \\
      $D^+_s \rightarrow \text{Extra} ~ \eta X$ & 6.0 & 0.0 & 0.0 & 1.6 & 1.6 & 7.2 & 1.6 & 1.6 & 7.2 \\
      $D^+_s \rightarrow \bar{s} X ~ (\text{no}~ \eta, \eta')$ & 2.1 & 1.0 & 0.0 & 0.0 & 0.0 & 0.0 & 1.0 & 0.0 & 0.0 \\
      $D^+_s \rightarrow \bar{s} X, X \rightarrow \eta$ & 1.0 & 0.5 & 0.0 & 0.3 & 0.3 & 1.2 & 0.8 & 0.3 & 1.2 \\
      $D^+_s \rightarrow \bar{s} X, X \rightarrow \eta'$ & 1.0 & 0.5 & 0.0 & 1.0 & 1.0 & 1.3 & 1.5 & 1.0 & 1.3 \\
      $D^+_s \rightarrow K^{0}_{S}K^{0}_{S} (K^{0}_{L}K^{0}_{L}) X$ & 3.3 & 3.3 & 0.0 & 0.0 & 0.0 & 0.0 & 3.3 & 0.0 & 0.0 \\
      $D^+_s \rightarrow K^{0}_{S}K^{+} (K^{0}_{L}K^{+}) X$ & 11.4 & 0.0 & 0.0 & 0.0 & 0.0 & 0.0 & 0.0 & 0.0 & 0.0 \\
      $D^+_s \rightarrow K^{0}_{S}K^{-} (K^{0}_{L}K^{-}) X$ & 3.7 & 7.5 & 0.0 & 0.0 & 0.0 & 0.0 & 7.5 & 0.0 & 0.0 \\
      $D^+_s \rightarrow K^{+}K^{-} (-\phi) X $ & 7.9 & 7.9 & 0.0 & 0.0 & 0.0 & 0.0 & 7.9 & 0.0 & 0.0 \\
      $D^+_s \rightarrow K^{+}K^{+} X$ & 0.1 & 0.0 & 0.1 & 0.0 & 0.0 & 0.0 & 0.0 & 0.1 & 0.0 \\
      $D^+_s \rightarrow K^{-}K^{-} X$ & 0.03 & 0.1 & 0.0 & 0.0 & 0.0 & 0.0 & 0.1 & 0.0 & 0.0 \\
      $D^+_s \rightarrow K^{0}_{S}K^{0}_{L} (-\phi) X$ & 0.0 & 0.0 & 0.0 & 0.0 & 0.0 & 0.0 & 0.0 & 0.0 & 0.0 \\
      $D^+_s \rightarrow e^{+} (\mu^{+}) X$ & 10.7 & -10.7 & 0.0 & 0.0 & 0.0 & 0.0 & -10.7 & 0.0 & 0.0 \\
      $D^+_s \rightarrow \tau^{+} \nu$ & 5.6 & 0.0 & 0.0 & 4.1 & 0.8 & 2.9 & 4.1 & 0.8 & 2.9 \\
      $D^+_s \rightarrow \mu^{+} \nu$ & 0.6 & 0.0 & 0.0 & 0.0 & 0.0 & 0.0 & 0.0 & 0.0 & 0.0 \\
      $D^+_s \rightarrow \text{Other~Annihilation}$ & 21.5 & 21.5 & 0.0 & 0.0 & 0.0 & 0.0 & 21.5 & 0.0 & 0.0 \\\hline
      Minimum Yields    &  &  &  &  &  &  & 96.2  & 20.5 & 46.8  \\
      Observed Yields   &  &  &  &  &  &  & 119.3 & 43.2 & 123.4 \\
      Additional Yields &  &  &  &  &  &  & 23.0  & 22.7 & 76.7  \\
           
      \hline \hline
    \end{tabular}
  \end{center}
\end{table*}      

For the 21.5\% yield of $D_s \rightarrow \text{Other~Annihilation}$
decays, we know nothing about the pion content other than that there
will be one $\pi^+$ to conserve charge. One might reasonably expect
that a substantial fraction of the 1/4 of the $D_s$ decays containing
an additional $\pi^+\pi^-$ pair would be in the ``Other Annihilation''
decays. As for the additional $\pi^0$ in 3/4 of the decays, that can
appear any place, e.g., as converting a charge-conserving $\pi^+$ into
a $\rho^+$. They will probably appear disproportionally in the 
``Other Annihilation'' decays, as these start (in our table) with
fewer particles.

The inclusive $\omega$ yield, $D_s \rightarrow \omega X$, of
$6.1\pm1.4\%$, is substantial. While $\omega$ has an $s\bar{s}$
component, it is {\it very} small, so it is unlikely that very much of
the $\omega$ yield comes from the $s\bar{s}$ component of 
$D^+_s \rightarrow s\bar{s} X$. At quark level, this is 
$D^+_s \rightarrow s\bar{s}u\bar{d}$, and a decay 
$D^+_s \rightarrow \pi^+ \eta \omega$ is quite possible. A decay 
$D^+_s \rightarrow \pi^+ \eta' \omega$, from energy considerations, is
just barely possible. From the decay 
$D^+_s \rightarrow s\bar{s}\bar{s}$, $\omega$ could come from 
$D^+_s \rightarrow K^+ \eta \omega$ (barely), but not from 
$D^+_s \rightarrow K^+ \eta' \omega$. 
From $D^+_s \rightarrow \bar{s} X$, it can come from 
$D^+_s \rightarrow K^+ \omega X$, with lots of phase space. And from 
``Other Annihilation'', there are lots of possibilities. In summary,
with the data we now have in hand, we can not say much about the
origin of the 6\% $\omega$ yield. A search for $D^+_s$ {\it exclusive}
decays will be reported in a separate paper. 
(We should note that our inclusive $\omega$ measurement came 
towards the end of the work described here, and so was {\it not}
included in the retuning of the Monte Carlo decay table. CLEO's $D_s$
Monte Carlo decay table produces far fewer $\omega$'s than the 6\% we
observe.)

In summary, we report several measurements of $D^+_s$
inclusive decays with significantly better precision than current
world averages.

We gratefully acknowledge the effort of the CESR staff
in providing us with excellent luminosity and running conditions.
D.~Cronin-Hennessy and A.~Ryd thank the A.P.~Sloan Foundation.
This work was supported by the National Science Foundation,
the U.S. Department of Energy,
the Natural Sciences and Engineering Research Council of Canada, and
the U.K. Science and Technology Facilities Council.

\end{document}